\RequirePackage{fix-cm}
\documentclass[twocolumn]{svjour3}          
\smartqed  
\usepackage{amsmath,graphicx,makecell}
\usepackage{enumitem}
\newlength\savedwidth
\newcommand\whline{\noalign{\global\savedwidth\arrayrulewidth
                            \global\arrayrulewidth 1.5pt}%
                   \hline
                   \noalign{\global\arrayrulewidth\savedwidth}}
\newlength\savewidth
\newcommand\shline{\noalign{\global\savewidth\arrayrulewidth
                            \global\arrayrulewidth 1.0pt}%
                   \hline
                   \noalign{\global\arrayrulewidth\savewidth}}
\usepackage{graphicx}
\usepackage{upgreek}
\usepackage{multirow}
\usepackage{multicol}
\usepackage{array}
\usepackage{cite}
\usepackage{color}

\begin{document}

\title{A Survey on Synchrophasor Data Quality and Cybersecurity Challenges, and Evaluation of their Interdependencies}
\author{Aditya Sundararajan, Tanwir Khan, Amir Moghadasi, Arif I. Sarwat}
 \institute{A. Sundararajan, T. Khan, A. Moghadasi, A.I. Sarwat\\
 EC 3920, 10555 W Flagler St, Miami, FL, USA\\
 Email: [asund005,tkhan016,amogh004,asarwat]@fiu.edu
}

\maketitle
\begin{abstract}
Synchrophasor devices guarantee situation awareness for real-time monitoring and operational visibility of the smart grid. With their widespread implementation, significant challenges have emerged, especially in communication, data quality and cybersecurity. The existing literature treats these challenges as separate problems, when in reality, they have a complex interplay. This paper conducts a comprehensive review of quality and cybersecurity challenges for synchrophasors, and identifies the interdependencies between them. It also summarizes different methods used to evaluate the dependency and surveys how quality checking methods can be used to detect potential cyber-attacks. In doing so, this paper serves as a starting point for researchers entering the fields of synchrophasor data analytics and security.
\keywords{Synchrophasors, data quality, cybersecurity, methodologies.}
\end{abstract}

\vspace{-0.5cm}
\section{Introduction}
\label{intro}
Smart grid has complex dependencies between physical and cyber realms~\cite{mqtt,theft15,theft18,theft16}. This has been demonstrated by recent attacks on smart grid (Table \ref{tab:attacks})~\cite{dagle,corsi,larsson,berizzi,gomes,imai}.
These attacks exploited a limited visibility of the system and inadequate support from reliability coordinators~\cite{theft17,theft22,bhatt,nerc1,zima,theft12,ashton,naspi1,gomes2,theft19,theft20,theft21}.
Wide-area measurement systems (WAMS) increase the situation awareness (SA) for operators~\cite{sutar,narendra,novosel}. WAMS devices that are part of the wide area monitoring, protection, automation and control include phasor measurement units (PMUs) at transmission, frequency disturbance recorders (FDRs) at low-voltage distribution and micro-PMUs ($\mu-$PMUs) for distributed renewables, called synchrophasors~\cite{gentz,reddy,amare,theft13,theft14,negash,rahman,nazari,chouhan,micropmu}.

Significant challenges to the implementation of synchrophasors have emerged in communication, data quality and cybersecurity. The existing communication infrastructure is slow, expensive and inflexible. To leverage SA and support timeliness, adequate quality checking methods must be in-place at the phasor data concentrators (PDCs) which aggregate and process raw data and flag corrupt data. Due to their ubiquity, synchrophasors have an increased attack surface.
The applications and challenges of synchrophasors are well-researched~\cite{sexauer,singh,sanchez,tushar,lauby,mohanta}. However, the challenges of data quality and cybersecurity are considered one independent of the other, when in reality, they are interdependent~\cite{pmudqnaspi,christo,zhu,yangbo,nadia,veda,sabha,danielson2013analysis,huyi,kdjones,huang,huang1,gpsspoof,morris,cyberlikeli,cyberthreats,hlin,beasley,tiago,energies,auth01,auth02,auth03,attack01,attack02,doesecchal,pmuchal,re10}.
Further, the literature does not leverage the knowledge of one challenge to address the other. For example, studying the changes to data quality can be key to potentially identify an underlying attack vector or an unexploited vulnerability.

\begin{figure}[t!]
\centering
\includegraphics[height=3in, width=3.25in]{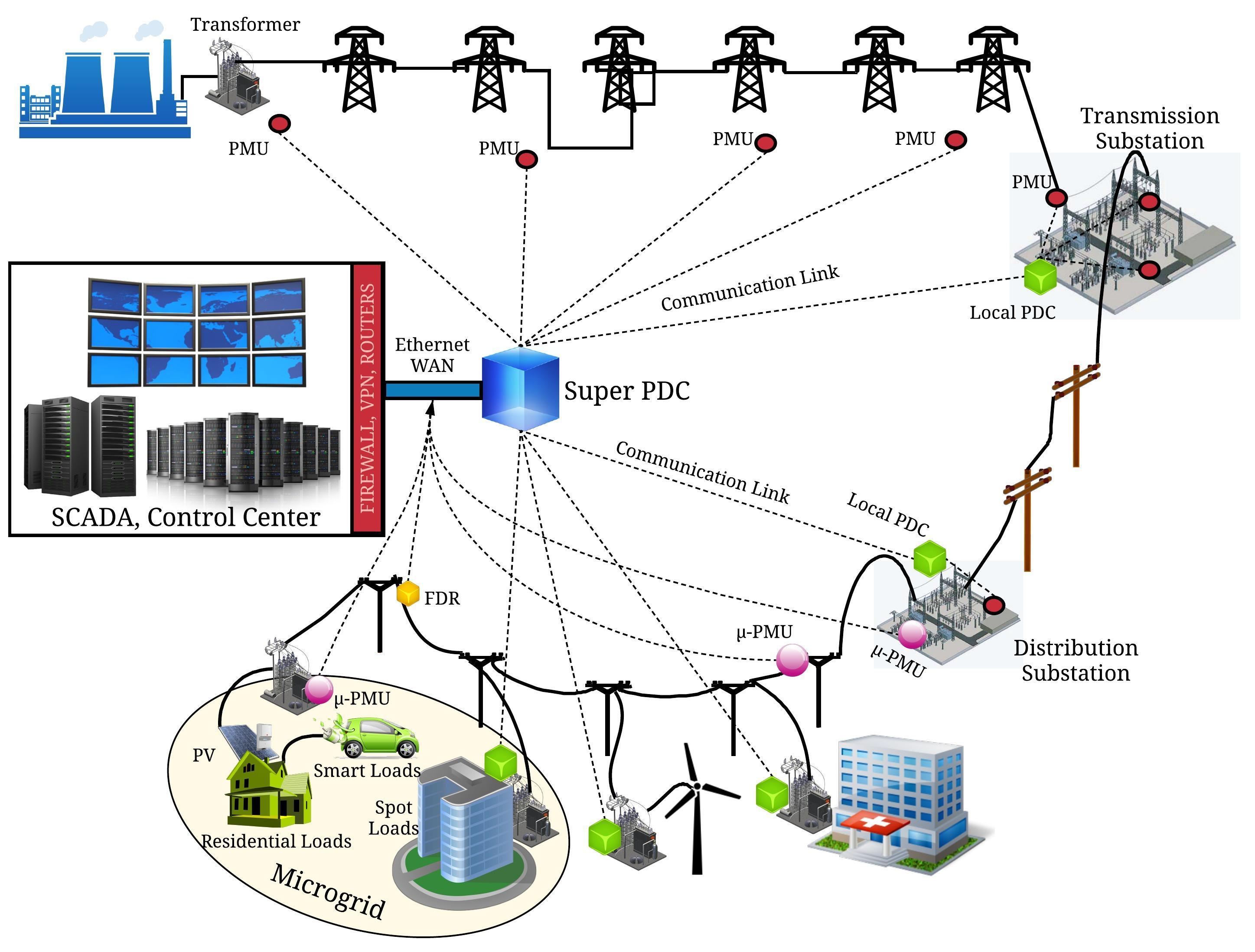}
\caption{Layout of smart grid WAMS comprising PMUs, $\mu-$PMUs, FDRs and PDCs}
\label{fig:one}
\end{figure}

The main contributions of this paper are: \textbf{(1)} maps the dependencies between data quality and cybersecurity challenges of synchrophasors; \textbf{(2)} reviews the methods to evaluate the challenges; and \textbf{(3)} surveys how data quality checking methods can leverage their observations to detect issues related to security. The paper also provides a high-level overview of synchrophasors, their standards, key applications, and challenges~\cite{johnson,martin,ali,retty}. It is key to know that poor quality can be due to device errors or communication challenges like congestion and packet collision. Similarly, all cyber-attacks do not impact the data, although reduction in quality is one of the biggest observable consequences of a successful attack. A layout of WAMS comprising synchrophasors is shown in Fig. \ref{fig:one}. This paper explores the challenges for PMUs at transmission and FDRs at distribution level.


\begin{figure}[b!]
\centering
\includegraphics[height=2.5in, width=3.25in]{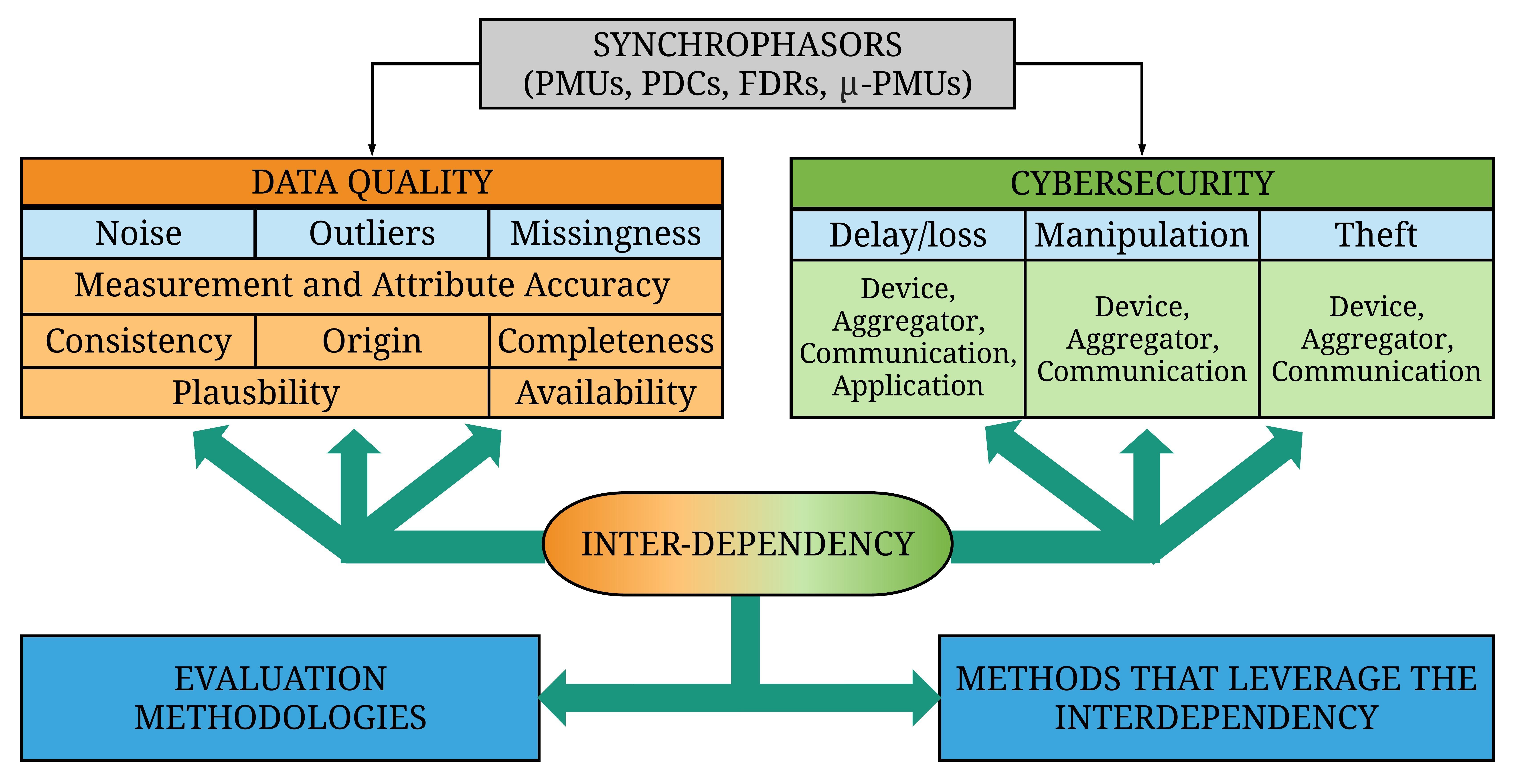}
\caption{The proposed structure of this survey paper}
\label{fig:layout}
\end{figure}

This survey paper considers data quality and cybersecurity as \textit{challenges}, where each has different \textit{issues}. Issues are the ways in which the particular challenge manifests when observed. Figure \ref{fig:layout} maps the challenges to their corresponding issues. The challenge of quality manifests in three ways: noise, outliers and missingness. Noise can be due to logical inconsistencies in data values or attributes while outliers result from poor integrity and origination. Missing data is a direct consequence of poor completeness and availability. Accuracy is impacted by noise, outliers as well as missingness while plausibility is a characteristic impacted by noise and outliers. These characteristics are discussed in Section \ref{sec:dq}. Cybersecurity manifests as delay/loss, manipulation or theft. While a delay/loss corresponds to a packet delay or drop due to congestion, timeout, buffer fullness or an intentional attack that affects availability, manipulation deals with attacks that alter the information, thereby impacting integrity. Theft captures attacks which compromise the confidentiality of data such as snooping, spoofing or espionage. These attacks occur at different levels of the synchrophasor hierarchy: \textit{device} corresponds to the edge devices like PMUs, FDRs, or $\mu-$PMUs, while \textit{aggregator} implies Local PDCs or SuperPDCs. \textit{Communication} refers to the synchrophasor network while \textit{Application} contains the different power system applications that use synchrophasor data.

The rest of the paper is organized as follows. Section \ref{sec:pmu} summarizes the architecture, major applications and key challenges of synchrophasors. The characteristics are described in Section \ref{sec:challenges}, and their interdependencies mapped in Section \ref{sec:inter}. While evaluation methods for data quality and cybersecurity are discussed in Section \ref{subsec:evaluation}, Section \ref{subsec:methods} surveys methods which use data quality characteristics to detect potential cyber-attacks. Section \ref{sec:conc} highlights future directions of research in synchrophasor data analytics and cybersecurity.

\vspace{-0.5cm}
\section{Architecture, Applications, Challenges}
\label{sec:pmu}
Synchrophasors can be standalone devices with dedicated purposes, or be a part of a larger system like the substations, depending on various functional and operational requirements. With increased penetration of renewables and smart loads, synchrophasors are used at distribution transformers and points of common coupling to study frequency disturbances and harmonics.
The architecture of synchrophasor devices are summarized at the device and network levels below.

\textbf{PMU device:}
It comprises Current Transformers (CTs) and Potential Transformers (PTs) that measure current and voltage magnitudes which are then converted to digital data, a microprocessor module that compiles these values, computes phasors, and synchronizes them with the Coordinated Universal Time (UTC) standard reference used by global positioning system (GPS) receivers that acquire a time-lag based on the atomic clock of GPS satellites~\cite{premerlani,ieee2,sutar,phadke,rihan}. They measure local frequency and its rate of change, and can record individual phase voltage and current along with harmonics, negative and zero sequence values~\cite{eval4}.
The information paints a dynamic picture of the grid at a given time. PMUs and PDCs transmit measured data as frames~\cite{ieee3}.
A $16$-bit cyclic redundancy check ensures data integrity. PDCs equipped with logging functionality use comma separated values or transient data exchange for data logs, and Common Format for Event Data Exchange for event logs~\cite{kezunovic,ieee4}. The data transfer rate of PMUs, which determine the message processing delays and network latencies, depend greatly on the timing requirements of applications.

\textbf{PMU network:} If there are multiple PMUs in a substation, Local PDCs aggregate site-level data and then transmit to a SuperPDC.
PDCs conduct various data quality checks and set flags according to the issues encountered, log performance, validate, transform, scale and normalize data, and convert between protocols~\cite{ieee1}.
There is typically a direct interface between PDC and the utility's SCADA or energy management system. PDCs can be deployed as standalone devices or integrated with other systems in the grid.

\textbf{FDR device:}
The Oak Ridge National Laboratory and the University of Tennessee Knoxville have been leading the FNET/GridEye project since 2004. FDRs have been installed and managed to capture dynamic behaviors of the grid. Although FDRs are essentially PMUs, they are connected at $120 V$, and hence incur lower installation costs than traditional PMUs do~\cite{chai}. FDRs are largely deployed at renewable integration zones of the grid, and measure nearly $1,440$ samples per second with a hardware accuracy of $\pm 0.5 mHz$ while PMUs measure between $10$ and $240$ samples per second and use GPS receivers that have $1\mu$s accuracy for synchronization~\cite{egypt,liu,stenbakken,wang}.
Given the availability of an extensive discussion of the architecture by the author of~\cite{fdrdev,fdr3rd}, it is beyond the scope of this paper.

\textbf{FDR network:}
FDRs use the internet to send data directly to the central servers for analytics and can provide information on transients, load shedding, breaker reclosing and the switching operations of capacitor banks and load tap changers~\cite{wang}. Unlike PMUs, FDRs can be installed at buildings and offices.

\textbf{Synchrophasor standards:} Multiple standards exist for PMU data measurement, transfer and communication, proposed by IEEE, the National Institute of Standards \& Technology (NIST), the North American Electric Reliability Commission (NERC) and the International Electrotechnical Commission (IEC)~\cite{ieee5,ieee6,ieee7,firouzi,mackiewicz,iec1,nistir,cip}. Due to multiple specifications and guidelines, there are possible contradictions in recommendations~\cite{mehta,johnson,martin,ali,retty}. A North American SynchroPhasor Initiative (NASPI) report in early 2016 identified the need for standardizing definitions related to synchrophasor data quality and availability by establishing the PMU applications requirements task force (PARTF)~\cite{naspi16}. IEEE standard C37.X deals with WAMS, specifically PMUs~\cite{ieee1,ieee8,ieee9}. These standards are summarized in Table \ref{tab:standards} with their core contributions highlighted. A more comprehensive review of the synchrophasor standards is documented in \cite{synchstd}.

\begin{table*}[h]
\renewcommand{\arraystretch}{1.3}
\caption{Various standards and guidelines for Synchrophasors}
\label{tab:standards}
\centering
\begin{tabular}{>{\centering\arraybackslash}c >{\centering\arraybackslash}p{3cm} >{\centering\arraybackslash}p{12cm}}
\whline
\textbf{Body} & \textbf{Standard} & \textbf{Core Contribution}\\
\shline
\multirow{ 12}{*}{IEEE} & 1344-1995 & Original parameter definitions for synchrophasors\\\cline{2-3}
& C37.118-2005 & Improved message formats, inclusion of time quality, Total Vector Error (TVE)\\\cline{2-3}
& C37.239-2010 & PMU/PDC event logging\\\cline{2-3}
& 1711-2010 & Serial SCADA Protection Protocol for substation serial link cybersecurity\\\cline{2-3}
& C37.118.1-2011 & PMU measurement provisions, performance requirements\\\cline{2-3}
& C37.118.2-2011 & Synchrophasor data transfer requirements\\\cline{2-3}
& C37.238-2011 & Common profile for applying Precision Time Protocol (PTP) using Ethernet\\\cline{2-3}
& C37.242-2013 & Synchronization, calibration, testing and installation of PMUs for PC\\\cline{2-3}
& C37.244-2013 & PDC functions and requirements for PC and monitoring\\\cline{2-3}
& C37.111-2013 & PMU/PDC data logging using COMFEDE\\\cline{2-3}
& 1686-2013 & Procuring, installing and commissioning IED cybersecurity\\\cline{2-3}
& C37.240-2014 & Sound engineering practices for high cybersecurity of substation APC\\\hline
\multirow{ 4}{*}{IEC} & 61850 & Interoperable and adaptable architectures to substation automation\\\cline{2-3}
& 61850-90-5 & Requirements for data exchange between PMUs, PDCs, PCs and control center\\\cline{2-3}
& 62351-1,2 & Security threats and vulnerabilities in smart grid devices\\\cline{2-3}
& 62351-6 & Prescribes digital signature using asymmetric cryptography for sending PMU data\\\hline
NERC & CIP 002-009 & Series of standards to ensure enterprise, field and personnel security\\\hline
NIST & NISTIR 7628 & Provides guidelines for smart grid cybersecurity (including WAMS)\\
\whline
\end{tabular}
\end{table*}

\textbf{Applications:} Synchrophasors streamline security, reliability and stability of power systems. They have online and offline applications~\cite{smartcity}. Online applications of PMUs include enhancing real-time SA, analyzing faults and disturbances, detecting and appraising oscillations and harmonics that impact power quality, and improving accuracy and reducing computational time of state estimation. Offline applications include congestion management, providing effective protection schemes, benchmarking, system restoration, overload monitoring and dynamic rating, validating the network model of SCADA, and improving overall power quality~\cite{novosel,iotsg,appmag}. Real-time (online) applications of FDRs include frequency monitoring interface integrated with command and control centers in the future for power system health diagnosis to prevent cascading failures, and event trigger module that detects and notifies the mismatch between generation and load caused by frequency variations. Offline applications include event visualization that renders the data read from the even data files~\cite{zhangfnet}.

\textbf{Challenges:} One of the major drawbacks of synchrophasors is the lack of transmission protocol, which makes them vulnerable to spoofing attacks~\cite{gentz}. The existing architecture is not scalable since it entails an initially high investment. NASPI's research initiative task force (RITT) emphasizes optimal placement as a significant challenge but also one dependent on the nature of applications the utility intends to use them for~\cite{naspi1}. The literature has multiple models including but not limited to genetic algorithm, simulated annealing, Tabu search, Madtharad’s method, particle swarm optimization, artificial neural networks, binary search and binary integer programming to address this challenge~\cite{reddy,amare,negash,rahman,nazari,chouhan,yuill,manousakis}.
More recently, managing and analyzing large volumes of synchrophasor data has become increasingly challenging. Lack of standardized data management solutions for smart grid has only made this problem more challenging. The ubiquitous presence of these devices has expanded their attack surface, making them vulnerable to different types of attacks. These two challenges are elaborated in the following section since they percolate to applications that directly operate upon the streaming data subject to minimal processing owing to timeliness requirements.

\vspace{-0.6cm}
\section{Data Quality and Cybersecurity Challenges in Synchrophasors}
\label{sec:challenges}
Due to their wide-ranging communication and automation capabilities, the challenges of synchrophasor data quality and cybersecurity have gained prominence.

\begin{table*}[h]
\renewcommand{\arraystretch}{1.3}
\caption{Summary of existing research in Synchrophasor Data Quality Challenges and Solutions}
\label{tab:dqchal}
\centering
\begin{tabular}{>{\centering\arraybackslash}p{3cm} >{\centering\arraybackslash}p{6cm} >{\centering\arraybackslash}p{7cm}}
\whline
\textbf{Attribute} & \textbf{Challenges} & \textbf{Solutions}\\
\shline
Completeness (Device, aggregator)~\cite{huyi,pmudqnaspi,wuh} & \vspace{-5mm} \begin{itemize}
\item PMU/PDC device damage
\item Faulty PMU-PDC communication
\item Network error
\item Database storage error
\item Data missing for failing to comply with latency and QoS requirements
\end{itemize} & \vspace{-5mm} \begin{itemize}
\item Acquiring better management techniques
\item Use of TCP protocol to re-transmit the lost data packets at the cost of timeliness
\item Adjusting the synchrophasor frame rate by increasing the wait time at PDC
\end{itemize} \\\hline
Measurement Accuracy (Device)~\cite{pmudqnaspi,christo,terzija,wuh,cokkinides,stenbakken1} & \vspace{-5mm} \begin{itemize}
\item Expected signal differs from measured signal due to harmonic interference
\item Introduction of noise to data
\end{itemize} & \vspace{-5mm} \begin{itemize}
\item Improving phase error using filtering techniques
\item NIST calibration per Standard C37.118-2005
\item Omnidirectional antennas
\item Context-based reconstruction of missing data
\item Network Time Protocol (NTP), e-Loran and Chip Scale Atomic Clock (CSAC)
\end{itemize}
\\\hline
Attribute Accuracy (Device, aggregator, communication)~\cite{pmudqnaspi,zhu} & \vspace{-5mm} \begin{itemize}
\item Measured vs actual timestamp discrepancy
due to satellite timing error; disagreement between encoded and actual location of PMU
\end{itemize} & \vspace{-5mm} \begin{itemize}
\item Development of linear state estimation tools
\item Avoiding timestamp discrepancy by modifying real-time clock element
\item Using OMP-based identification, BB algorithm to solve PMU location discrepancies
\end{itemize}\\\hline
Plausibility and Availability (Communication, application)~\cite{veda,sabha,kdjones,heiko,maab,danielson2013analysis} & \vspace{-5mm} \begin{itemize}
\item Impact of measurement system on individual data points
\item Data inaccessibility due to high network latency or device failure.
\item Delayed data arrival due to increased routing traffic 
\end{itemize} & \vspace{-5mm} \begin{itemize}
\item Use of Electrical Data Recorder (EDR) tools for capturing high-rate time series data, data storage and analysis
\item A more lenient time limit could be set for non-critical application usage.
\item Latency-aware application design
\end{itemize} \\\hline
Origination (Device, aggregator, communication, application)~\cite{pmudqnaspi,castello,dua,huang1} & \vspace{-5mm} \begin{itemize}
\item Poor standard interpretation, implementation
\item Misalignment, erroneous compression
\item Latency, loss of communication nodes
\item Data corruption due to delivery time of PDC exceeding permissible slot
\item Network unavailability to process incoming data streams
\end{itemize} & \vspace{-5mm} \begin{itemize}
\item Redundancy in communication by using wireless and wired connections
\item Lagrange interpolating polynomial method
\item Data substitution, imputation, interpolation and extrapolation
\item Stochastic forecasting with Prediction Error Minimization (PEM)
\end{itemize}\\\hline
Logical Consistency (Aggregator, communication)~\cite{pmudqnaspi} & \vspace{-5mm} \begin{itemize}
\item Data transmitted contains no headers
\item Sampling rate of data changed at PMU without being adjusted at PDC
\item Data duplication while processing
\item Data from different PMUs with incorrect timestamps
\end{itemize} & \vspace{-5mm} \begin{itemize}
\item Logical consistency can be ensured by maintaining the PMU registries and data protocols
\end{itemize}\\
\whline
\end{tabular}
\end{table*}

\vspace{-0.5cm}
\subsection{Data Quality Challenges}
\label{sec:dq}
NERC's real-time tools best practices task force (RTBPTF) and NASPI's PARTF impose requirements to ensure synchrophasor data quality~\cite{pmudqnaspi,pmudqnerc}. Data quality can be contextualized in different ways, depending on the needs of the concerned domain. For instance, data quality requirements of a smart meter recording energy consumption might differ from those of a net meter at a solar photovoltaic (PV) power plant. NASPI contextualizes synchrophasor data quality to determine ``fitness of use" in terms of accuracy and lineage for static data points; lineage, completeness and logical consistency for static datasets; and availability, timeliness and origination for streams of data points~\cite{pmudqnaspi}.

There could be different causes for poor data quality: a) Device: poor calibration of device, biases due to CT, PT; erroneous filter design, poor synchronization of timing measurements, and issues due to measurement channel; b) Communication: latency exceeding stipulated limits, network congestion, signal interferences and failure of communication nodes; c) Aggregator: Data transformation resulting in errors, delayed arrival of packets dropped due to time-limit exceeding, and unwanted duplication or corruption of data during computations; and d) Application: storage and maintenance issues, insufficient training size, erroneous manipulations to the data and poor association of context. Although data quality requirements vary with applications, they have been extensively documented~\cite{synchstd,huang,pmudqnaspi,pmudqnerc}.
The existing literature on synchrophasor data quality is summarized in Table \ref{tab:dqchal}.

\noindent \textbf{Completeness:} focuses on the gaps between different values, accounts for missing values~\cite{pmudqnaspi}. The attributes of completeness defined at device and aggregator-levels are:
\textit{gap rate}- number of gaps in data per unit time; \textit{mean gap size}- mean of the length of known gaps; and \textit{largest known gap}- length of the largest known gap among the different gaps. While completeness is impacted by device malfunction, packet drops and communication link failure, the literature does not recognize the possibility of an attack behind such causes.\\
\noindent \textbf{Accuracy:} can be of the value or attribute, primarily measured in Total Vector Error (TVE), which according to IEEE standard C37.118, is the vector difference between the measured and expected phasor value (magnitude, angle and frequency). Accuracy is categorized into that of: \textit{data values}- impacted by factors like the difference between expected and observed signals or the introduction of noise to the data within the synchrophasor; and \textit{data attributes}- affected by factors like accuracy of the measured timestamp, agreement between encoded and actual location coordinates of the device, and alignment of the location recorded in the power system topology with its actual location~\cite{nadia}.

\noindent \textbf{Plausibility and Availability:} Measurement specifiers are the attributes of data which describe whether the process of measuring some phenomenon of the power system (observed value) and calculating its value (expected value) are documented effectively in terms of standard units to a given precision and are within a stated confidence interval~\cite{nadia,sabha}. These specifiers have decisive sub-attributes influencing the qualitative value of data: data representing the measurement of quality or condition of the grid, and data represented in the form of SI units up to 3 decimal places with a confidence interval included.

Network availability plays an important role in streaming data~\cite{danielson2013analysis}, and in-turn affects data availability. In case of high network latency, the incoming data streams from different synchrophasors get delayed or lost, causing applications to perceive them as missing or incomplete. Hence, network availability can be considered an indirect attribute affecting quality. This can be mitigated if the overlying applications are programmed to account for the delays, or if a more lenient waiting time limit is set. However, the second solution is dependent on the kind of applications the synchrophasors cater to. The latency requirements for synchrophasors recommended by the standards are very stringent.

\noindent \textbf{Origination:} is the source from which the data is measured. Its trustworthiness is associated with the background and source. Its attributes are: \textit{point of origin:} the class of device from where the data originated (Measurement (M) or Performance (P) for PMUs), the standard followed by the device, and any data manipulation or standardization techniques through which the data passes~\cite{eval5,pmudqnaspi}; \textit{coverage}: physical location of the device based on its geospatial or electrical topology location~\cite{zhu,yangbo}; and \textit{transformation} applied to the data at the device, aggregator or application level.

\begin{table*}[h!]
\renewcommand{\arraystretch}{1.3}
\caption{Summary of the Recent Cyberattacks on Smart Grid Impacting Data Quality}
\label{tab:attacks}
\centering
\begin{tabular}{>{\centering\arraybackslash}p{3cm} >{\centering\arraybackslash}p{3.5cm} >{\centering\arraybackslash}p{3cm} >{\centering\arraybackslash}p{2.5cm} >{\centering\arraybackslash}p{3cm}}
\whline
\textbf{Source of Attack (Year)} & \textbf{Target of Attack} & \textbf{Data Quality Characteristic Impacted} & \textbf{Cybersecurity Characteristic Impacted} & \textbf{Attack Specifics}\\
\shline
Vulnerability in network firewall (2001) & California ISO (CAISO) web servers & Consistency, accuracy & Integrity & Poor security configuration during planned maintenance\\\hline
Stuxnet worm (2010) & Programmable Logic Controllers (PLCs) at SCADA & Accuracy, consistency, plausibility & Integrity, Availability & Exploits zero-day vulnerabilities of PLCs\\\hline
BlackEnergy (2011) & Human-Machine Interface of utility grid control systems & Plausibility, origin, accuracy, consistency & Confidentiality, Integrity, Availability & General Electric's HMI targeted\\\hline
Remote Access Trojan; watering-hole attack (2014) & ICS/SCADA & Plausibility, origin, accuracy, consistency & Confidentiality, Integrity, Availability & Conducted by Dragonfly, Energetic Bear\\\hline
Trojan.Laziok reconnaissance malware (2015) & Energy companies & Origin, plausibility & Confidentiality & Gathered information from compromised devices\\\hline
BlackEnergy3 (2015) & Ukrainian grid control center & Plausibility, origin, accuracy, consistency & Confidentiality, Integrity, Availability & Lack of SA left 220,000+ customers without power\\\hline
WannaCry ransomware cryptoworm (2017) & Computers running MS Windows Operating System & Availability, origin & Availability & Used EternalBlue, a vulnerability in older Windows systems\\
\whline
\end{tabular}
\end{table*}

\noindent \textbf{Consistency:} determines how agreeable the data is with the overall structure of its type. Incompatibility of attributes in terms of measurement rates or header labeling between datasets results in outliers, leading to an inconsistent result from an application. The attributes of consistency are:
\textit{Header frame consistency:} consistency of the Header frame of the device. This could be categorized into: persistence of PMU Header that states whether the PMU header structure is consistent over time, and persistence of PDC Header that states whether the PDC header structure is consistent over time;
\textit{Data frame consistency:} consistency of Data frames of the device. This could be categorized into: persistence of PMU Data frame that states whether the PMU data frame structure is consistent over time, and persistence of the PDC Data frame that states whether the PDC data frame structure is consistent over time;
\textit{order consistency of Data frames:} whether the order in which the Data frames are recorded is consistent in the device; \textit{consistency in compliance to standards} recommended for PMU and all the devices associated with it; and \textit{consistency of reporting rate:} whether data reporting rate is consistent across all devices.

Emerging research in this area has lately focused on determining solutions for ensuring data quality. These solutions include using omnidirectional antennas to improve GPS availability, context-aware determination of missing data streams using accurate timing information, Network Time Protocol (NTP) and associated chip scale atomic clocks (CSACs) as backups for synchronization when GPS fails, imputation, interpolation and extrapolation, stochastic forecasting with prediction error minimization (PEM) and data substitution~\cite{huang}.

\noindent \textbf{Evaluation of Quality:} Methods to evaluate quality is discussed in Section \ref{subsec:evaluation}. The approach for performance evaluation is to first study the impacts of device calibration and network conditions on quality, then examine how poor quality reduces the application performance~\cite{pmudqnaspi}. Two effective methods are proposed to evaluate the impact of quality on performance: \textit{benchmarking} that tests an application multiple times with numerous erroneous datasets in contrast to those with no known errors, and \textit{standardization} that documents, for each application, the level of tolerable errors.

\vspace{-0.6cm}
\subsection{Cybersecurity Challenges}
\label{sec:cs}

\begin{table*}[h]
\renewcommand{\arraystretch}{1.3}
\caption{Summary of Existing Research in Synchrophasor Cybersecurity Challenges and Solutions}
\label{tab:attackchal}
\centering
\begin{tabular}{>{\centering\arraybackslash}p{3cm} >{\centering\arraybackslash}p{4cm} >{\centering\arraybackslash}p{9cm}}
\whline
\textbf{Level} & \textbf{Challenges} & \textbf{Solutions}\\
\shline
Device, Aggregator~\cite{synchstd,flexpdc,wuh,cip,gpsspoof,morris} & \vspace{-5mm} \begin{itemize}
\item Device damage
\item Device calibration tampering
\item Forging PMU data
\item GPS spoofing
\end{itemize} & \vspace{-5mm} \begin{itemize}
\item Multi-alteration technique to trace adversary in event of GPS spoofing
\item Visible GPS satellite prediction
\item Anomaly between expected and measured GPS signals
\item Using SSL/TLS or IPSec to encrypt data before transmission
\item Using state estimation technique to mitigate device calibration and tampering
\item Rigorous penetration testing prior to installation
\end{itemize}\\\hline
Communication~\cite{cokkinides,stenbakken1,zhushi,heiko,maab,cyberthreats,energies,auth01,attack02} & \vspace{-5mm} \begin{itemize}
\item Denial of Service
\item Man-in-the-Middle
\item False Data Injection
\item Snooping attack
\item Delay attack
\end{itemize} & \vspace{-5mm} \begin{itemize}
\item Airgapping PMU network
\item Filtering routers, disabling IP broadcasts, applying security patches, disabling unused ports
\item Server authentication by clients before establishing connection
\item Use of time-series state estimation
\item Cryptographic methods like AES, DES
\item Mutual authentication
\item Cyber trust model with blockchains
\item NASPInet hub-spoke model
\item Optimal key generation and distribution
\end{itemize}\\\hline
Application~\cite{castello,dua,auth02,auth03,attack01} & \vspace{-5mm} \begin{itemize}
\item Phishing and social engineering
\item APT and insider threats
\item Replay attacks
\end{itemize} & \vspace{-5mm} \begin{itemize}
\item Authentication, Authorization and Accounting (AAA)
\item Use of secure data transfer protocol to prevent replay attack
\end{itemize}\\
\whline
\end{tabular}
\end{table*}

Synchrophasors cater to applications like state estimation, contingency analysis and optimal power flow that need real-time high-resolution data measurement, communication and analytics~\cite{inlsec}. Therefore, a successful attack on these devices might cause erroneous SA or cascading failures~\cite{synchrodqup,cyberlikeli}. Yet, many industrial organizations do not consider synchrophasors as critical cyber assets. Recent cyberattacks on the smart grid (Table~\ref{tab:attacks}) mostly used powerful malware like worms, viruses or Trojan horse, but a few attacks like the one on the Pacific Gas \& Electric transmission substation relied on physical means. These attacks jeopardized not just the availability of power but also that of control data (information).
Cybersecurity of synchrophasors are categorized into: 1) Device, Aggregator, 2) Communication, and 3) Control center application.

\noindent \textbf{Device, Aggregator:} NASPI Network (NASPInet) is logically capable of integrating WAMS across multiple geographically distant organizations using phasor gateways (PGWs). The attacks at this level compromise data integrity, targeting devices from individual PMUs to PDCs, SuperPDCs or even PGWs. Some attacks include: 1) tampering the signal measurement units of devices through interference, 2) illicitly changing the calibration of devices to report erroneous readings, 3) forging data to reflect wrong measurements, and 4) GPS spoofing by broadcasting fabricated signals to the receiver to yield erroneous synchronization of phasors computed, modifying satellite position, or replaying legitimate GPS signals at later timestamps~\cite{gpsspoof}. 

GPS spoofing can be mitigated by enabling the receiver to predict visible GPS satellites at a given position and time instant and use the Coarse/Acquisition (C/A) code from those satellites. Another strategy compares the measured GPS signal to the estimated signal and computes the anomaly error which must have an accuracy of $\leq 40ns$ for nearly $95\%$ of the values according to IEEE C37.118~\cite{zhu}.
Synchrophasors must be subject to rigorous testing before installation. Some methods include port scans, device security feature robustness, network congestion testing, denial of service testing, network traffic sniffing and disclosure testing~\cite{morris}. These tests should be periodically conducted by certified white hat penetration testers after installation. Regular patches and configuration updates must be made down to the end-device level.

\begin{table*}[!t]
\renewcommand{\arraystretch}{1.3}
\caption{Summary of the Interdependency between Quality and Cybersecurity Challenges}
\label{tab:interdep}
\centering
\begin{tabular}{>{\centering\arraybackslash}p{2cm} >{\centering\arraybackslash}p{4cm} >{\centering\arraybackslash}p{3.5cm} >{\centering\arraybackslash}p{3.5cm} >{\centering\arraybackslash}p{2cm}}
\whline
\textbf{Level} & \textbf{Quality Attribute} & \textbf{Quality Issue} & \textbf{Cyber-Attack Observed} & \textbf{Security Attribute Impacted}\\
\shline
Device & Completeness, accuracy, plausibility & Synchronization signal loss, measurement signal loss, missing data & GPS spoofing, replay, device tamper, changing device calibration, FDI & Integrity \\\hline
Aggregator & Origin, consistency, plausibility & Corrupted data, anomalies, outliers & FDI, Tampering, Buffer overflow, MITM & Confidentiality, Integrity\\\hline
Communication & Availability, origin, consistency & Anomalies, outliers, inconsistent, out-of-order data & DoS, MITM, FDI, snooping, replay, delay & Confidentiality, Integrity, availability\\\hline
Application & Origin, availability, consistency, completeness, accuracy & DoS, delay, APT, FDI, theft/fraud, insider attack & corrupted data, missingness, anomalies, outliers & Confidentiality, Integrity, Availability\\
\whline
\end{tabular}
\end{table*}

\noindent \textbf{Communication:} Synchrophasors support bidirectional communication channels, where data measurements flow from devices to the control center while control signals flow the other way. The vulnerabilities of the protocols used by the devices also contribute to the overall security. Attacks on communication channels compromise integrity, availability and confidentiality. Some attacks include: \textit{1) Denial of Service (DoS)} by overwhelming PMUs, PDCs or other aggregation devices higher in the hierarchy with bogus frames so that legitimate frames are lost, delayed, denied or dropped, \textit{2) Man-in-the-Middle (MITM)} attacks by a malicious entity posing itself as PDC (to PMU) or PGW (to PDC) and sending malicious commands that causes PMUs/PDCs to behave in an abnormal manner that triggers failures, \textit{3) False Data Injection (FDI)} by intercepting frames over the channel, altering or replacing them with malicious information that then gets propagated to higher levels of the WAMS, \textit{4) Snooping} by the attacker eavesdropping on the channel for incoming or outgoing frames, typically not modifying or stealing but just capturing a copy of that information for packet replay or espionage, and \textit{5) Delay} caused by compromising communication routers that deliberately induce latencies in propagation to critically affect the grid's SA.

Many authentication and authorization algorithms are proposed to secure synchrophasor data over communication channels~\cite{cyberthreats}. These methods range from conventional encryption methods to cyber trust. Due to the ubiquity and widespread range of these devices, key distribution and management becomes a problem. Mutual authentication is also proposed to account for trust~\cite{energies}. Decentralized, blockchain-based trust acquisition is being considered too. The publish-subscribe hub-spoke architecture proposed by NASPInet supports dynamic sharing of device data to alleviate shortcomings of the communication medium like delays and latencies. Standards like IEC 61850-90-5 recommend trusted key distribution center to generate and distribute keys that meet system requirements~\cite{auth02,auth03,attack01,attack02}.

\noindent \textbf{Application:} Despite being protected by enterprise security tools for intrusion detection and prevention, virtualization, segmentation, authentication, authorization and access control, cyberattacks still proliferate~\cite{doesecchal,pmuchal}. It is understood that any successful attack at the other two levels perpetrated in a manner undetectable by the enterprise security systems can pose a significant threat. The attacks at this level are the most dangerous, since crucial power system applications use data from WAMS to conduct analysis to address reliability, power quality, network topology, and faults. An adverse impact on these calculations could compromise the ``self-healing" nature of the grid. More recent solutions include game theory, machine learning, proactive data visualization, and Defense-in-depth~\cite{theft22,human}.

\vspace{-0.25cm}
\subsubsection{Evaluation of Security}
Works have tested the resilience of PMUs and PDCs against different attacks. The authors in~\cite{cybersec1} conducted penetration testing of a synchrophasor network in IEEE 68-bus system to map vulnerabilities against the Common Vulnerabilities Exposure (CVE) database. Potential corrective measures to ensure the security of PMUs and PDCs is proposed~\cite{cybersec2}. Considering the security at substation and information levels, the authors provide a wide range of tools to mitigate breaches at both fronts. A multilayered architecture at the substation is proposed where different levels of data abstraction is provided between PMUs and external environment, supplemented by firewalls, User Datagram Protocol (UDP) Secure for communication over untrusted networks, and remote access using Secure Shell (SSH).

\vspace{-0.6cm}
\section{The Data Quality-Cybersecurity Dependency}
\label{sec:inter}

The severity of an attack can be understood from the extent of its impacts on the targeted system. With the smart grid encouraging interoperability between devices, information, applications, and protocols, a transparent and direct information exchange is now feasible. This also means that if information in one of the interconnected systems is infected, it is bound to propagate to other systems upon exchange, affecting the whole network. Synchrophasor devices harbor such vulnerabilities, as summarized in Section \ref{sec:cs}. However, to mitigate cyberattacks on interconnected systems, the relationship between devices and data must be known.

\begin{table*}[!h]
\renewcommand{\arraystretch}{1.3}
\caption{Summary of Evaluation Methods for Quality (DQ) and Cybersecurity (CS) Issues}
\label{tab:interdepeval}
\centering
\begin{tabular}{>{\centering\arraybackslash}p{3cm} >{\centering\arraybackslash}p{6cm} >{\centering\arraybackslash}p{7cm}}
\whline
\textbf{Issue} & \textbf{Challenge} & \textbf{Evaluation Methods}\\
\shline
Noise (DQ) & Consistency, accuracy & \vspace{-5mm} \begin{itemize}
\item Cable configuration, testing, validation
\item Specifying confidence interval, precision, TVE, ROCOF for measurements
\item Evaluating instrumentation channels
\begin{itemize}
\item Model-based correction
\item State estimation-based error filtering
\end{itemize}
\item Presistence in Data/Header frames
\item Standards compliance
\end{itemize}\\\hline
Outlier (DQ) & Consistency, origin, accuracy & \vspace{-5mm} \begin{itemize}
\item Standardization, benchmarking
\item Enhancing endpoints with switches, routers
\item Specifying device model, coverage and content
\end{itemize}\\\hline
Missingness (DQ) & Completeness, availability, accuracy & \vspace{-5mm} \begin{itemize}
\item Dedicated communication channels
\item Enhancing endpoints with switches, routers
\end{itemize}\\\hline
Delay/loss (CS) & All levels & \vspace{-5mm} \begin{itemize}
\item Regular penetration testing of all levels
\item Link-level encryption, selective encryption
\item Dedicated communication channels
\item Data redundancy for fault tolerance
\end{itemize}\\\hline
Manipulation (CS) & Device, Aggregator, Communication & \vspace{-5mm} \begin{itemize}
\item Regular penetration testing of all levels
\item Link-level encryption, selective encryption
\item Data abstraction, multi-layered architecture
\item Data redundancy for fault tolerance
\item Augmenting ID/IPS, firewalls, ACLs, VPNs
\end{itemize}\\\hline
Theft (CS) & Device, Aggregator, Communication & \vspace{-5mm} \begin{itemize}
\item Regular penetration testing of all levels
\item Data abstraction, multi-layered architecture
\item Data redundancy for fault tolerance
\item Augmenting firewalls, ACLs, VPNs
\end{itemize}\\
\whline
\end{tabular}
\end{table*}

Table \ref{tab:interdep} summarizes key interdependencies between the two challenges. There is a tight coupling between data quality and cyber-attacks, implying it is wise to study synchrophasor cybersecurity by accounting for the impacts on quality. In most attacks, plausibility, completeness, accuracy and consistency are primarily impacted~\cite{anurag,kzhu}. In Section \ref{subsec:evaluation}, specific evaluation methods for quantifying this relationship are reviewed. Section \ref{subsec:methods} looks at how data quality characteristics can be used as markers to detect potential cyber-attacks within the context of synchrophasors. Results from these subsections are summarized in Tables \ref{tab:interdepeval} and \ref{tab:interdepmethods}, respectively.

\vspace{-0.5cm}
\subsection{Interdependency Evaluation Methods}
\label{subsec:evaluation}
Next to communications, cybersecurity was found to impact the design and installation costs for synchrophasors~\cite{eval2}. This is because they are critical to the mission-support systems of the grid. Different practical ways for utilities to mitigate quality issues like accuracy, timeliness and consistency are also identified. Some methods include employing dedicated communication channels between PMUs and PDCs, encrypting PMU data before communication, and enhancing communication endpoints using firewalls and routers. The report, however, does not delve into the details of how such methods could impact latency (and hence, timeliness) and availability of the data.

Given different manufacturers of devices, there will be differences in measurement and calibration quality despite adhering to the standards. The varying application requirements cause differences in application-level PMU performance, of which data quality is a major one. The static and dynamic PMU testing efforts of the Performance and Standards Task Team (PSTT) of NASPI and the PMU performance characterization are briefly summarized in~\cite{eval6}. In it, the different steady-state tests performed on magnitude, phase and frequency evaluate their conformance to accuracy requirements, which is an important attribute of data quality and is a direct target of many cyberattacks. Given the impact of instrumentation channels on the quality, they have been well-characterized and evaluated for impacts on accuracy in the literature. The errors induced by them could be rectified through model-based correction algorithms and state estimation based error filtering. Some other avenues where data quality could be evaluated include the cable configurations, testing and validating the devices to ensure accurate, consistent performance and interoperability at all levels~\cite{pstt1,pstt2}. Although not explicit, these works hint at the improvement in the resilience of synchrophasor devices against potentially malicious activities by accounting for proper testing methods to characterize and evaluate the different sources of errors prior to deployment that might contribute to poor quality.

\begin{table*}[!t]
\renewcommand{\arraystretch}{1.3}
\caption{Summary Showing How Quality Can Help Identify Cybersecurity Issues}
\label{tab:interdepmethods}
\centering
\begin{tabular}{>{\centering\arraybackslash}p{3cm} >{\centering\arraybackslash}p{3cm} >{\centering\arraybackslash}p{4cm} >{\centering\arraybackslash}p{6cm}}
\whline
\textbf{Cyber-attack} & \textbf{Quality Affected} & \textbf{Quality Check Looks For} & \textbf{Mitigation Methods Using Quality}\\
\shline
Device tampering (delay/loss, theft) \newline \cite{exploringchallenges,synchsec} & Completeness, plausibility, accuracy, consistency, origination & Large gap sizes, inaccurate readings, ping fail  & \vspace{-5mm} \begin{itemize}
\item Statistical substitution: regression, imputation, interpolation
\item Intelligent substitution: neural networks, logistic regression, optimization
\item Securing the physical devices
\end{itemize}\\\hline
Spoofing PMU data (manipulation) \newline \cite{fdi2,dataspoof,dataspoof2} & Consistency, accuracy, plausibility & Unexpected values, errors, mismatch with SCADA values, redundant timestamp, out-of-order packet arrival & \vspace{-5mm} \begin{itemize}
\item Monitoring line impedances for anomalies
\item Divergence and miscorrelation between SCADA and PMU data
\end{itemize}\\\hline
GPS Spoofing (manipulation, delay/loss) \newline \cite{gpsspoof,eval3,gpsspoofing1,gpsspoofing2} & Consistency, origination, plausibility & Inaccurate timing value, TVE $> 1\%$, packets arrive out-of-order & \vspace{-5mm} \begin{itemize}
\item Using multiple synchronization sources or telecommunications
\item Anti-spoofing checking methods at receivers
\item Internal holdover oscillators as backups for providing accurate timing signals
\item Spoofing match algorithm with Golden Search for lighter computation
\end{itemize}\\\hline
Denial of Service (delay/loss) \newline \cite{synchsec} & Completeness, accuracy, consistency & Congestion at PDCs/network, delayed arrival of packets, dropped packets, inability to reach suspected device & \vspace{-5mm} \begin{itemize}
\item Augmenting PDCs with inline blocking tools
\item Employ port hardening and disable IP broadcasts
\item Use high bandwidth communications (expensive)
\end{itemize}\\\hline
Man-in-the-Middle (delay/loss, manipulation, theft) \newline \cite{synchsec} & Origination, accuracy, availability, consistency & Mismatch between obtained and expected value, abnormal delay in packet arrival & \vspace{-5mm} \begin{itemize}
\item Mutual authentication, message authentication codes
\item Digital certificates with active management of CRLs
\end{itemize}\\\hline
False Data Injection (manipulation, theft) \newline ~\cite{fdi1,fdi3,fdi4,fdi5,fdi6} & Plausibility, consistency, accuracy, origination & Mismatch with SCADA values, unexpected values, spatio-temporal outliers & \vspace{-5mm} \begin{itemize}
\item Spatio-temporal correlations, density-based local outlier factoring
\item Monitor line impedance for anomalies
\item Random time-hopping of packets
\item Divergence and miscorrelation between SCADA and PMU data
\end{itemize}\\\hline
Snooping, Sniffing (theft) \newline \cite{synchsec,beasley} & Plausibility, origin & No observable changes- additional analysis needed & \vspace{-5mm} \begin{itemize}
\item Using secure gateway/VPN communication
\item Employing TLS/SSL, SSH, lightweight selective encryption
\end{itemize}\\\hline
Delay (delay/loss) \newline \cite{synchsec,beasley} & Completeness, consistency, availability, accuracy & Observable patterns in gaps, slow arrival of packets & \vspace{-5mm} \begin{itemize}
\item Statistical and intelligent substitutions
\item Redundant measurement devices on the same line
\end{itemize}\\\hline
APT, Insider threat (delay/loss, theft, manipulation) \cite{human} & Accuracy, consistency, origin, plausibility & No observable changes- additional analysis needed & \vspace{-5mm} \begin{itemize}
\item Defense-in-depth
\item Machine learning, advanced data analytics
\end{itemize}\\
\whline
\end{tabular}
\end{table*}

Final conclusions can be gathered from~\cite{eval7}. The report by the Pacific Northwest National Laboratory (PNNL) analyzes existing synchrophasor networks in terms of their communication and information-level interoperability, security and performance. It concluded that latency is a key issue for the future synchrophasor designs which is expected to compound latency due to PDC functionality. It also emphasized that substations generally did not employ redundancy; there is little consistency in adoption of security methods for synchrophasor networks. Some tools include link-level encryption, virtual private networks (VPNs), ID/IPS, firewalls and access control lists (ACLs). Further, existing data quality checking methods locate a compromise in integrity by identifying faulted data values (due to measurement errors, communication delays or external events) but not due to result of device tampering, MITM, spoofing or FDI. Since both faults and attacks have the same impact on quality, it is important to differentiate the two causes while checking for the attributes such as accuracy, consistency and timeliness.

To summarize, the following measures can be used as metrics to quantify data quality: TVE, errors in magnitude, phase, frequency and ROCOF, harmonics and noise for measurement accuracy; comparison between measured and expected results, confidence interval and precision for measurement specifiers; temporal, geospatial and topological accuracy for attribute accuracy; device model specifications, geospatial and topological coordinates, coverage and content for origination; persistence in Header and Data frames, standards compliance, reporting rate and order for logical consistency; and gap rate, gap size and largest known gap for completeness. Benchmarking and standardization are two methods that can be used to evaluate data quality. Similarly, cybersecurity can be quantified by conducting extensive penetration testing of the synchrophasor networks integrated into benchmarked IEEE bus systems for different types of attacks (DoS, MITM, FDI, spoofing, probing, cache poisoning) and discovering potential vulnerabilities that could be exploited. While doing so, it would be important to also repeat the evaluation of the quality attributes using the above metrics and explore how they are impacted due to the specific attacks, and whether they violate the industry standards requirements specified for different applications.

\vspace{-0.5cm}
\subsection{Addressing Cyber-attacks Using Quality Issues}
\label{subsec:methods}

It can be seen from Table \ref{tab:interdepmethods} that successful cyberattacks compromise synchrophasor data quality since the security requirements are violated~\cite{cybersecreq}. Given synchrophasors use TCP/UDP on the transport layer for their communications, attacks typically possible on TCP/IP stack like DoS, MITM, packet replay or spoofing are possible in synchrophasor domains as well.

Physical attacks like device tampering causes loss or incurs theft of critical information, easily observed through large gaps sizes, poor accuracy in obtained values and unreliable origin. The lost data is typically handled through substitution, either statistical or intelligent~\cite{exploringchallenges,synchsec}. The best way to prevent physical attacks like cable disconnects, direct damage to device, etc. is by ensuring the devices are isolated from external weather and human elements.

Spoofing synchrophasor data is achievable through polynomial fitting or data mirroring techniques. Such attacks impact quality that manifests as outliers or noise. Several methods have been proposed to counter these attacks: intra-PMU and inter-PMU correlations to determine the relationship between PMU parameters and across PMUs in a locality, respectively; machine learning techniques like Support Vector Machines (SVMs) and more~\cite{fdi2,dataspoof,dataspoof2}.

GPS spoofing exploits publicly available civilian GPS signals using air or cable to produce signals that initially align with the original, but slowly start increasing the power to drown the authentic signal and thereby compromising the receiver~\cite{gpsspoof,eval3}. By introducing measurement errors in the time synchronization, the attacks induce changes in data consistency and plausibility which can be used as markers to identify the likelihood of the attack~\cite{gpsspoofing1,gpsspoofing2,yuder}.

In a successful DoS where multiple synchrophasor devices get compromised, packet delay or loss is observed. This impact in quality can serve a clue to the onset of DoS-style attacks. Typical solutions involve augmenting inline blocking tools, high bandwidth connections, disabling IP broadcasts and port hardening.

MITM is possible in synchrophasors where the attacker acts as a legitimate PDC to the PMUs and vice-versa, thereby intercepting and/or modifying all messages exchanged. This is noticed by quality checking methods in the form of poor accuracy and consistency in values between what was sent by PMU and what was received by PDC. It can be avoided by having the devices employ mutual authentication and a digital certificate mechanism with an actively managed Certificate Revocation Lists (CRLs) and certificate authorities~\cite{synchsec,beasley}.

FDI impacts the consistency, accuracy and plausibility of the data. The effects are typically observed as spatio-temporal outliers in the data. Quality checking methods check for this anomaly and may employ correlation across different timestamps to identify the corruption of data. FDI is one of the widely explored attacks on synchrophasor domain, with solutions like determining the mismatch between the values obtained from PMUs and that observed in SCADA, monitoring the line impedances which get affected when data is manipulated, and using density-based Local Outlier Filter (LOF) analysis~\cite{fdi1,fdi3,fdi4,fdi5,fdi6}.

Sometimes, attackers simply capture the packets flowing in a channel with an intent to listen. Such sniffing/snooping attacks have been conducted using WireShark to realize messages are exchanged in plaintext. This attack is difficult to detect using data quality checking methods since most often, no quality characteristic is impacted as the attackers do not affect the data actively. However, technologies like VPN, encryption of selective messages (to reduce the overall process overhead), or Transport Layer Security (TLS)/Secure Socket Layer (SSL), Secure Shell (SSH) can be used to mitigate them. While TLS has been shown to be susceptible to poisoning attacks and VPN to side channel attacks, careful network design can account for them~\cite{synchsec,snooping}.

With the increased frequency of campaign efforts and nation-sponsored attacks against the grid, synchro-phasors could be lucrative targets for sophisticated attacks like advanced persistent threats (APTs), social engineering, watering-hole attacks and malware-based intrusions~\cite{tong,flau,lanchao,jzhang,hastings}. While these attacks scale beyond specific devices in the synchrophasor hierarchy, the quality checking methods alone would not be sufficient~\cite{human}. The use of defense-in-depth model augmented with stakeholder interactions, awareness and training, and intelligent solutions like machine learning for attack data classification and/or event prediction, root-cause analysis of observed events, developing evolving defense topographies using moving target defense, and advanced visualization techniques for efficient cognition of events would play a critical role.

The key takeaway from this section is that impacts on data quality can provide strong markers for an underlying cyber-attack. Noise, outliers and missing values are all commonly observed issues which quality checking methods may be programmed to detect, analyze and base decisions on. Certain sophisticated attacks like APTs, insider threats, sniffing, and social engineering have indirect impacts on quality which a checking method may not be able to detect with enough confidence or precision. Additional solutions are required to mitigate such attacks in the synchrophasor domain. These solutions include statistical methods like divergence, correlation, regression and substitution; intelligent methods like neural networks and evolutionary algorithms for event classification and prediction, logistic regression for substitution; technologies like VPNs, firewalls, ID/IPS, anomaly detectors, selective encryption, port hardening, network isolation and use of TLS/SSL, SSH; and human-in-the-loop solutions like advanced visualization techniques, awareness and training, and stakeholder engagements. While the impacts on quality can also be due to underlying device or measurement errors, most of the works in the literature assume the data has been subject to delay/loss, manipulation or theft intentionally. This paves way for the recommendation that the upcoming research in this area must look at ways to differentiate the impacts on data quality due to attacks from errors.

\section{Future Directions of Research and Conclusion}
\label{sec:conc}
The future directions of research in the areas of synchrophasor data quality, cybersecurity and communications are multi-faceted. Addressing data quality challenges must begin with a strong push to the adoption of industry-wide, vendor-agnostic data management, processing and storage standards for smart grid. Most recent cyber-attacks were successful due to the difference in speed of cognition of the information generated by automated vulnerability detection tools and the speed with which the machine data is created (called cognitive gap)~\cite{human}. The design of synchrophasor devices are also expected to improve in the future~\cite{smartcity}. Keeping in mind the quality challenges, an improvement to PDC design called flexible integrated synchrophasor system (FIPS) was proposed to minimize issues in quality and communication, and tackles specific tasks of PDC such as data alignment, employs cryptographic methods to ensure confidential exchange of data without jeopardizing integrity, and establishes relevance to the NASPInet~\cite{flexpdc}. To ensure device and application-level interoperability, development of technical standards and conformance testing rules is expected. Further, the emergence of distribution-level $\mu-$PMUs will evoke the need for developing measurement, communication, quality and security standards. Further, with the deployment of distributed renewable sources, electric and autonomous vehicles, energy storage and transactive energy, there is a strong impetus for enhancing technologies behind monitoring and control, of which synchrophasors will play a major role~\cite{eval2}.

To conclude, while existing research has focused on the synchrophasor challenges of quality and cybersecurity individually, their interdependency has largely been ignored. This paper makes one of the first attempts at highlighting the impacts of cyber-attacks on various quality attributes, thereby recommending that the future research on the design and development of security solutions should account for their impacts on quality as well, and that different quality characteristics can be used by quality checking methods to flag for potential cyber-attacks. Plausibility, completeness, accuracy and consistency are some of the attributes that are most adversely impacted by a majority of the attacks on synchrophasors. At the same time, not all cases of poor data quality imply a successful cyber-attack as the reason. Different metrics that could be used to quantify quality attributes were summarized, and the methods that help evaluate the impacts of quality and security on performance were also briefly highlighted. This paper serves as a starting point for researchers entering these areas as it summarizes and determines their interdependency and relevance to smart grid security.

\section*{Acknowledgments}
This material is based on work supported by the National Science Foundation Grant No. CNS-1553494 and the Department of Energy Grant No. 800006104. Any opinions, findings, and conclusions or recommendations expressed in this material are those of the authors and do not necessarily reflect the views of the NSF and DOE.

\end{document}